\documentclass[fleqn,usenatbib]{mnras}

\usepackage{newtxtext,newtxmath}
\usepackage[T1]{fontenc}
\usepackage{ae,aecompl}
\usepackage{algorithm}
\usepackage[noend]{algpseudocode}
\usepackage{hyperref}
\usepackage{color}
\usepackage{float}
\usepackage{placeins}
\usepackage{graphicx}	
\usepackage{amsmath}	
\usepackage{amssymb}	
\usepackage{caption}

\newcommand{\sub}[1]{_{\rm #1}}

\newcommand{\vimp}{v\sub{imp}}

\newcommand{\beq}{\begin{equation}}
\newcommand{\eeq}{\end{equation}}

\newcommand{\GoogleEarth}{{\tt Google Earth}}
\newcommand{\Vinales}{{Vi\~nales}}

\newcommand{\hll}[1]{#1}
\newcommand{\hl}[1]{\textcolor{black}{#1}}
\newcommand{\mytitle}{Can we predict the impact conditions of \hl{metre}-sized meteoroids?}

\newcommand{\aimp}{1.217}
\newcommand{\eimp}{0.391}
\newcommand{\eapprox}{0.4}
\newcommand{\qimp}{0.74}
\newcommand{\iimp}{11.47}
\newcommand{\iapprox}{11}
\newcommand{\Wimp}{132.28}
\newcommand{\wimp}{276.97}
\newcommand{\Pimp}{1.34}
\newcommand{\Tpimp}{2.81}

\newcommand{\lonimp}{-83.8037}
\newcommand{\latimp}{+22.8820}
\newcommand{\Arad}{178.9}
\newcommand{\hrad}{31.8} 
\newcommand{\vimpval}{16.9} 
        
\newcommand{\pinarhmax}{30} 
\newcommand{\pinarhmin}{27} 

\newcommand{\maxheight}{67} 
\newcommand{\maxbright}{22} 

\title[Can we predict Impact conditions]{\mytitle}

\author[Zuluaga et.al]{
Jorge I. Zuluaga$^{1}$\thanks{Corresponding author: jorge.zuluaga@udea.edu.co},
Pablo A. Cuartas-Restrepo$^{1}$,
Jonathan Ospina$^{2}$ and Mario Sucerquia$^{1}$
\\
$^{1}$Solar, Earth and Planetary Physics Group - SEAP,  Institute of Physics, University of Antioquia, calle 70 No. 52 - 21, Medell\'in (Colombia)
\\
$^{2}$Sociedad Antioque\~na de Astronom\'ia / CAMO group, Medell\'in (Colombia)
}

\date{Accepted XXX. Received YYY; in original form ZZZ}

\pubyear{}

\begin{document}
\label{firstpage}
\pagerange{\pageref{firstpage}--\pageref{lastpage}}
\maketitle

\begin{abstract}
\hl{Every year,} a few \hl{metre}-sized meteoroids impact the atmosphere of the Earth.  Most (if not all) of them are undetectable before the impact. Therefore, predicting where and how they will fall seems to be an impossible \hl{task}. In this letter we show compelling evidence that we can constrain in advance, the dynamical and geometrical conditions of an impact. For this purpose, we \hl{analyse} the well-documented case of the Chelyabinsk \hl{(Russia)} impact and the more recent and smaller \hl{\Vinales\  (Cuba)} event, whose conditions we estimate and provide here. After using the {\em Gravitational Ray Tracing} algorithm (GRT) to ``predict'' the impact conditions of the aforementioned events, we find that the speed, incoming direction and (marginally) the orbital elements \hl{of the corresponding meteoroids could be constrained} in advance, starting only on one hand, with the geographical location and time of the impact, and on the other hand, with the distribution in configuration space of Near Earth Objects (NEOs). Any improvement in our capability to predict or at least to constrain impact properties of medium-sized and large meteoroids, will help us to be better prepared for its potentially damaging effects.
\end{abstract}


\section{Introduction}
\label{sec:introduction}

Earth is impacted by hundreds to thousands of \hl{centimetre}-sized meteoroids each year. \hl{Metre}-sized objects are much less frequent, with just a few of them falling into the Earth every year \citep{Brown2013}.  Most of these events occur at very high altitudes and \hl{far from populated areas.} Still, they are  detectable by satellites \citep{Chapman1994, Brown2002} and infrasound detectors (see eg. \citealt{Silber2011, Moreno2018}). In a few cases, however, an impact of a \hl{metre}-sized object (releasing energies in the range of few to several to hundreds of ktons) may happen over populated areas, posing a real risk over infrastructure and population \citep{Popova2013, Rumpf2016}. 



In the case of large (several hundreds of meters) and well-knwon Near Earth Object (NEOs) whose impact probabilities are not negligible, we can predict in advance where and how the object could impact our planet \citep{Chapman2004, Chesley2005, Rumpf2016}. This is possible because the impactor have been previously observed and its orbital elements are very well-constrained.

The detection of \hl{metre}-sized NEOs is slow and difficult (see eg. \citealt{Boslough2015}).  Therefore, predicting where and how one of these \hl{unseen} objects will impact our planet at a given time, seems to be an impossible task. Proof of this are the \hl{well-known} impacts in Chelyabinsk \hl{(Russia)}, Benenitra (Madagascar) and recently in \hl{Vi\~nales (Cuba)} and the \hl{Bering Sea, (Russia)}, which \hl{were} not observed prior to the event.

In two recent works, \citet{Zuluaga2017p,Zuluaga2018} introduced a novel numerical technique, the Gravitational Ray Tracing (GRT), intended (among many potential applications) to  compute the probability that at a given time, certain geographical region of the Earth (or any other planetary body) be impacted by a \hl{meteoroid}.  More recently \citet{Zuluaga2019} applied GRT to study the impact of a small object against the moon, testing the technique for the first time in a different context and with a different aim for which it was originally devised.

In this letter we apply GRT (Section \ref{sec:GRT}) to study retrospectively the impact conditions (Section \ref{sec:conditions}) of the best documented atmospheric explosion \hl{to date, namely} the Chelyabinsk impact, \hl{hereafter the Chelyabinsk-Russia or C-R event} (Section \ref{sec:events}), and the \hl{more} recent, smaller, but still energetic event over \hl{\Vinales\ (Cuba), hereafter the \Vinales-Cuba or V-C event}.  Our aim here is to verify if the ``predictions'' that GRT could make about the incoming direction, speed and orbital element of the potential impactors at the place and time of those events (Section \ref{sec:tratm}), \hl{are close to} the observed conditions of the impacts.  \hl{In order to perform the comparison between the well-known C-R event with the barely-studied V-C impact}, we provide our own estimations of the impact conditions as obtained from public footage \hl{and} using the methods introduced in \citealt{Zuluaga2013} (Supplementary material).

\vspace{-0.2cm}

\section{Impact conditions}
\label{sec:conditions}

\hl{After encountering the Earth, the fate of a meteoroid depends on many different parameters, including its pre-entry size, mass and incoming direction \citep{Gritsevich2012,Lyytinen2016}.}  \hl{Still, } most \hl{(if not all)} \hl{metre}-sized objects, ie. diameters $1 \lesssim D \lesssim 50$ m, \hl{that have impacted the Earth in the last 30 years leaving a detectable fireball, have exploded at altitudes ranging 20-50 km}\footnote{\hl{For an updated list of reported fireballs in the CNEOS database see: \url{https://cneos.jpl.nasa.gov/fireballs/}}}\hl{, in accordance with most theoretical expectations \citep{Svetsov1995,Collins2005}}.  \hl{Ablation  and subsequent fragmentation}, happen when the atmospheric density increases and the aerodynamic pressure at the leading edge of the impactor surpass the strength of the material \citep{Hills1998}. 
Usually, the ``cascade'' breakup of the main body spreads  material of the meteoroid over a relatively large area of the planet surface. 

We call ``impact conditions'' to the set of bulk geometrical and dynamical properties of a meteoroid impact on the \hl{Earth's atmosphere.  These conditions include but are not restricted to the entry speed, incoming direction, and projected impact point.}  


We call {\it projected impact point} to the intersection between the geoid and a straight line tangent to the atmospheric trajectory of the meteoroid. Their geodetic coordinates are denoted as lon$_{\rm imp}$, lat$_{\rm imp}$. The point in the sky from which the meteoroid seems to come is known as the {\it radiant}. \hl{Here,} we parameterize the radiant in terms of the Azimuth ($A_{\rm rad}$) and elevation ($h_{\rm rad}$) of \hl{this direction} as observed from the projected impact point.

\hl{Since} the impact speed $\vimp$ \hl{of the meteoroid} is \hl{relatively large},  typically $\vimp\sim 12-24$ km/s, and its mass is \hl{much} larger than the mass of the atmosphere displaced during the \hl{ablation}, the trajectory of the \hl{meteor} can be assumed nearly rectilinear until it reaches the \hl{upper layers of the lower-atmosphere, ie. altitudes $\sim$ 20-30 km}.  \hl{Therefore, after determining} the \textit{projected impact point} and the \textit{radiant direction} \hl{we can} estimate the heliocentric orbital elements \hl{prior to impact}, namely $q$, $e$, $i$, $\Omega$ and $\omega$ (with $q$ the perihelion distance, $e$ the eccentricity, $i$ the orbital inclination, $\Omega$ the longitude of ascending node and $\omega$ the argument of perihelion). \hl{These elements are obtained by integrating the trajectory of the object, subject to the gravitational field of the Solar System, back to a reasonable time previous to impact (typically one orbital period before).}
\vspace{-0.3cm}
\section{Studied events}
\label{sec:events}

In recent years, the expansion and densification of urban areas \hl{plus} the availability of cheap electronic cameras, allowed that two large impact events be witnessed \hl{and registered by thousands to millions of casual observers.} 
The \hl{first one} was the Chelyabinsk-Russia (C-R) event (February 15, 2013)\hl{,  the largest reported fireball since the Tunguska-Russia impact}.  \hl{More recently a smaller, still energetic,  impact} was witnessed in \Vinales, Cuba (February 1, 2019), \hl{the V-C event}. Other relatively large events, such as one in Madagascar (July 27, 2018), had also thousands of witnesses, but much less available \hl{footage} and data. Multiple sources of information, including public footage, satellite imagery and \hl{recordings} from infrasound networks, has allowed us to determine with incredible detail (at least in the case of C-R impact) the impact conditions of these \hl{two} events. 

We provide in the \hl{supplementary material, full details of a detailed} estimation of the V-C event impact conditions, as obtained \hl{exclusively} from public footage, and applying the methods \hl{previously introduced} in \citealt{Zuluaga2013}. In Table \ref{tab:events} we show the impact conditions for the C-R and V-C events, as obtained from the available literature \citep{Zuluaga2013,Popova2013,Borovivcka2013} and according to our own estimations. 

\begin{table}
\centering
\begin{tabular}{lll}
\hline\hline
Property & \hl{C-R}$^1$ (2013) & \hl{V-C}$^2$ (2019)\\
\hline
\multicolumn{3}{c}{Atmospheric Trajectory}\\
\hline
                &  \citealt{Borovivcka2013} & This work \\
Date            & 2013/02/15 & 2018/02/01 \\
Time (UTC)      & 03:20:20 & 18:17:10 \\
lon$_{\rm imp}$ (deg)       & +59.8703* & \lonimp \\
lat$_{\rm imp}$ (deg)       & +55.0958* & \latimp \\
$A_{\rm rad}$ (deg)         & 103.5 & \Arad \\
$h_{\rm rad}$ (deg)         & 18.55 & \hrad \\
$\vimp$ (km/s)  & 19.03 & \vimpval \\
\hline
\multicolumn{3}{c}{Heliocentric Orbit}\\
\hline
                & \citealt{Borovivcka2013} & This work \\
$a$ (AU)        & 1.72 & \hl{\aimp} \\
$e$             & 0.571 & \hl{\eimp} \\
$q$ (AU)        & 0.738 & \hl{\qimp} \\
$i$ (deg)       & 4.98 & \hl{\iimp} \\
$\Omega$ (deg)  & 326.459 & \hl{\Wimp} \\
$\omega$ (deg)  & 107.67 & \hl{\wimp} \\
$P$ (years)     & 2.33 & \hl{\Pimp} \\ 
$T_p$**           & 2.73 & \hl{\Tpimp} \\ 
\hline\hline
\multicolumn{3}{l}{\hl{$^1$C-R, Chelyabinsk-Russia,} \hl{$^2$V-C, Vi\~nales-Cuba.}}\\ 
\multicolumn{3}{l}{*\citealt{Zuluaga2013}}\\
\multicolumn{3}{l}{
**Tisserand parameter, $T_p=1/a+2\cos i\sqrt{a(1-e^2)}$}
\end{tabular}
\caption{Impact conditions of the impact events studied in this work.\label{tab:events}}
\end{table}

\begin{figure*}
  \centering
  \vspace{0.2cm}
   \includegraphics[scale=0.38]{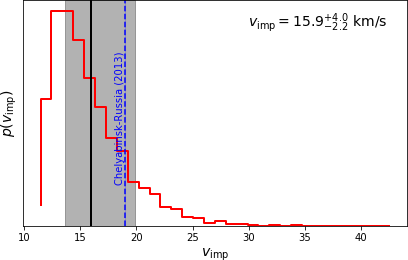}\hspace{0.5em}%
   \includegraphics[scale=0.38]{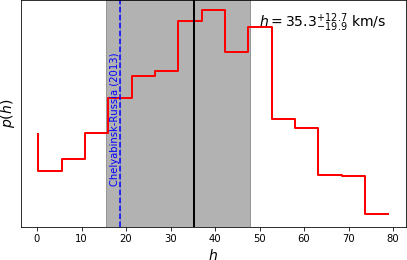}\hspace{0.5em}%
   \includegraphics[scale=0.38]{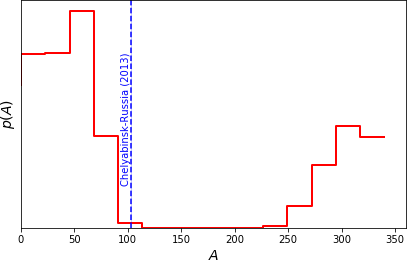}\hspace{0.5em}%
   \\
   \includegraphics[scale=0.38]{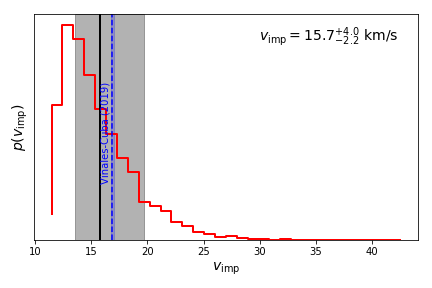}\hspace{0.5em}%
   \includegraphics[scale=0.38]{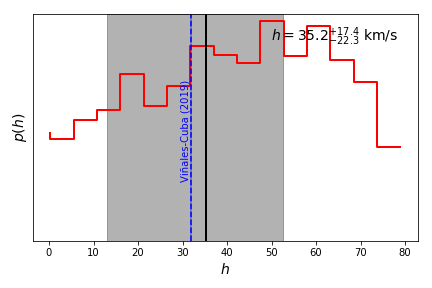}\hspace{0.5em}%
   \includegraphics[scale=0.38]{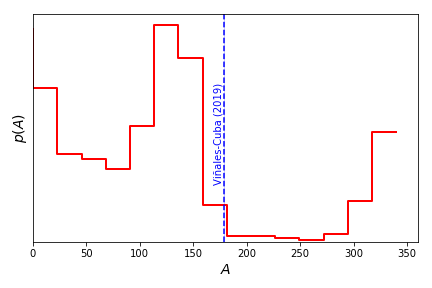}\hspace{0.5em}%
  \caption{Marginal probability distributions of impact velocity, elevation and azimuth for C-R (upper row) and V-C (lower row) events.}
\label{fig:marginal}
\end{figure*}
\vspace{-0.3cm}

\section{Gravitational Ray Tracing}
\label{sec:GRT}

Are the impact of meteoroids and their \hl{specific} conditions predictable?  As stated above, the small-size of most of the objects \hl{falling into the Earth each year}, prevents their early  detection, \hl{making} very improbable, if not \hl{entirely} impossible, to anticipate their place and conditions of arrival.  

\citet{Zuluaga2017p,Zuluaga2018}  (hereafter ZS2018) developed 
and tested a backward integration technique intended to compute the statistical properties of \hl{these} impacts. The technique was inspired in the ray tracing algorithms used in the film and game industries to render photorealistic images (see eg. \citealt{Comninos2010} and references there in), \hl{which also inspired the name of the technique, \textit{Gravitational Ray Tracing} (GRT)}.  

In GRT, we randomly generate N different impact velocities regularly spaced \hl{in the interval 11.1-43.6 km/s} (the Earth's escape velocity and \hl{the the biggest possible velocity achievable at Earth's orbit respectively, see ZS2018 for details}) and M random incoming directions following a blue-noise distribution to avoid aliasing sampling effects (see Section 2.1 in ZS2018).  Starting at a given impact site and a desired date and time, we integrate backwards the trajectory of these $N\times M$ test particles in the Solar System gravitational field.  The integration stops when test particles reach an asymptotic heliocentric orbit. 

The probability that a test particle, having impact conditions $(A,h,\vimp)$, correspond to a real meteoroid, is given by the so-called {\it ray probability}:

$$
P(A,h,\vimp;t) \propto  f(\theta\sub{apex},\lambda\sub{apex}) R(q,e,i,\Omega,\omega).
$$

Here $R(q,e,i,\Omega,\omega)$ is the number density of NEOs in the orbital elements space \hl{(see Sec. \ref{sec:discussion} for a discussion)}. In order to correct for the ``defocusing''  effect that the relative motion of the Earth has with respect to the NEOs population, we introduce a flux correcting factor $f$ (see Section 2.5 in ZS2018):

\begin{eqnarray}
f(u=\cos[90-\theta\sub{apex}];a,b)&=&
\left\{
\begin{array}{ll}
u^a & -1\leq u < 0; \\
u^b & 0\leq u \leq 1.
\end{array}
\right., 
\label{eq:Flux}
\end{eqnarray}

where $\theta_{\rm apex}$ and $\lambda_{\rm apex}$ are the polar and azimutal angle of the incoming direction with respect to the apex (direction of motion of the Earth). $a\approx 1$ and $b\approx 0.5$ are constants that we fit using the frequency as a function of apex angle of meteors in the CNEOS fireball reports database. The number density of NEOs around a point $\vec{x}\equiv(q,e,i,\Omega,\omega)$ in configuration space is estimated with:
\begin{eqnarray}
R(\vec x) &=& \sum_k W(||\vec{x}-\vec{x}_k||,\eta),
\label{eq:SmoothingFunction}
\end{eqnarray}
 
where $||\vec{x}-\vec{x}_k||$ is a generalized ``distance'' and $\eta$ a scale parameter.  $W(||\vec{x}-\vec{x}_k||,\eta)$ is called the {\it smoothing kernel}, and it is intended to soft the transition from a discrete to a continuous regime (see Eq. 7 in ZS2018).  \hl{The sum in $R$ include all the NEOs $k$ at a distance less than $2\eta$ of the point $x$.}

Distances in configuration space $||\vec{x}-\vec{x}_k||$ are computed using the \citet{Zappala1990} metric with the parametrization introduced by \citet{Rozek2011}:

\hll{
\beq
\label{eq:Z-metric}
\begin{array}{lll}
(D_Z/n_m a_m)^2 & = &
\frac{5}{4} (a-a_m)^2/a_m^2 \,+\, 
2 (e - e_k)^2 \,+\, \\ 
&  & \,+\, 2 (\sin i - \sin i_k)^2 \,+\,\\
&  & \,+\, 
10^{-4} (\Omega - \Omega_k)^2 \,+\,
10^{-4} (\varpi - \varpi_k)^2
\end{array} 
\eeq
}

with $a_m=(a+a_k)/2$ the average semi-major axis between, $n_m$ the corresponding orbital mean motion and $\varpi=\Omega+\omega$ the longitude of the perihelion.

\begin{figure*}
\centering
\includegraphics[scale=0.45]{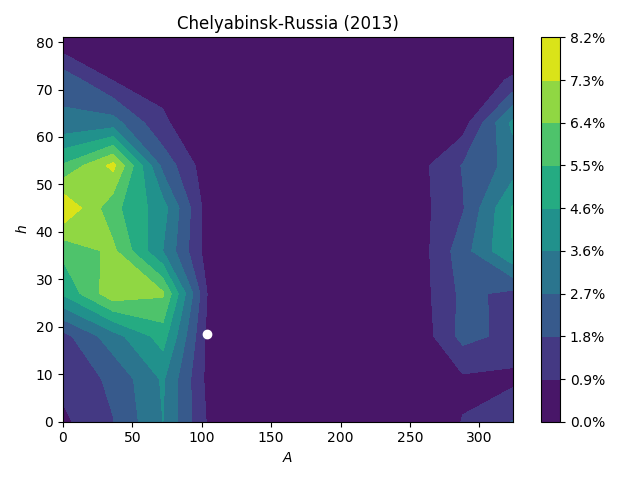}\hspace{2em}%
\includegraphics[scale=0.45]{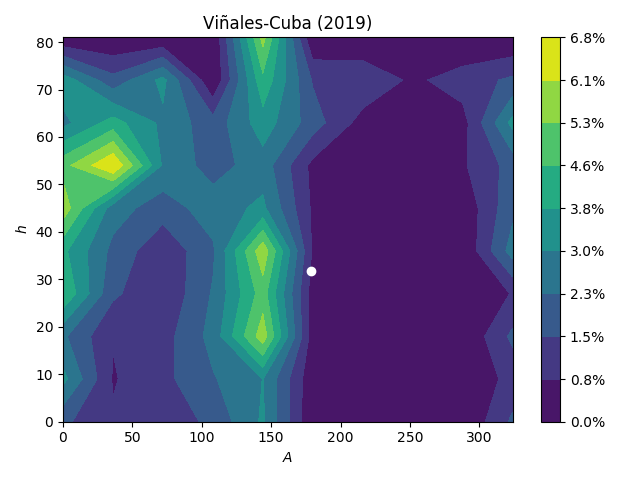}
\caption{The estimated radiant for the Chelyabinsk (left panel) and Cuba meteor events (right panel), and the most prone meteoroid radiants in colourmaps according to the GRT technique.}
\label{fig:Azh}
\end{figure*}

\subsection{Marginal Probabilities}

In ZS2018, we sum-up the ray probabilities at a given geographical location, to compute the (relative) probability that an object \hl{hit} that location instead of another one.  This allows us to create impact risk maps. But ray probabilities can be used in different ways. Thus, for instance, we can estimate the \hl{\it marginal probability distribution} of impact speeds, ie.  $p(\vimp)$, if we sum-up the probabilities \hl{for all rays} having impact speeds in the interval ${\cal V}:[\vimp, \vimp+\Delta\vimp]$:

$$
p(\vimp) \, \Delta\vimp\propto \sum_{\cal V} P(A_i,h_i,v\sub{imp,i};t),
$$

The same rationale can be applied for computing the marginal probability distribution of any impact condition (radiant elevation, radiant azimuth, asymptotic orbit semi-major axis, etc.) In Figure \ref{fig:marginal} we show the marginal probability distributions for several quantities, as computed \hl{with} GRT at the location and time of the C-R and V-C events.  \hl{A comparison with the true impact conditions is also shown.}

A similar approach can be \hl{used} to compute \hl{\it bivariate marginal probability distributions}.  Thus, for instance, the probability that the radiant elevation and Azimuth be in the solid angle $\Delta\Omega:[h_{\rm rad},h_{\rm rad}+\Delta h_{\rm rad}], [A_{\rm rad},A_{\rm rad}+\Delta A_{\rm rad}]$ can be estimated by:

\hl{
$$
p(A_{\rm rad},h_{\rm rad})\,\cos h_{\rm rad}\, \Delta h_{\rm rad} \, \Delta A_{\rm rad}\, \propto \sum_{\Delta\Omega} P(A_i,h_i,v\sub{imp,i};t),
$$
}

\hl{It should be noticed that in this case a geometrical correcting factor, namely $\cos h_{\rm rad}$, should be introduced to estimate the probability distribution function, since solid-angles at larger elevations are smaller than those near to the horizon.}

\vspace{-0.5cm}
\section{Results}
\label{sec:tratm}

Impact conditions (Figure \ref{fig:marginal}), for both the C-R and V-C events, are well within the statistical errors of the  \hl{GRT predictions}. The case of azimuth is especially interesting.  Although the observed value of this quantity, have low marginal probability in both events, GRT predicts \hl{correctly the octant} in the sky from which the impactors arrived. In the case of \hl{C-R event}, GRT predicted a north-east radiant, \hl{$A_{\rm rad}\sim 50-70^\circ$}. \hl{In the real impact,} the object appeared coming from the east, \hl{$A_{\rm rad}\sim 100^\circ$}. More interesting is the case of the V-C meteoroid.  \hl{Public observations reported to} the American Meteor Society, AMS\footnote{\url{https://fireball.amsmeteors.org/members/imo_view/event/2019/513}}, \hl{as well as early interpretation of the available footage}, suggested a southbound meteor direction. \hl{The theoretical predictions with GRT (see rightmost panel in the lower row  of Figure \ref{fig:marginal}) predicted that meteoroid should enter traveling from south to north (radiant in the south, $A_{\rm rad}\sim 140-160^\circ$), as was later confirmed by an independent analysis of satellite images.}


In Figure \ref{fig:Azh} we show \hl{contour maps of the bivariate marginal probability distribution} of azimuth and elevation. These maps reveal details absent in the univariate marginal distributions of Figure \ref{fig:marginal}.  \hl{Thus, for instance, according to CRT} at the date, time and location of the C-R event, almost no objects should come from azimuths between 100$^\circ$ and 270$^\circ$ (half of the sky), nor from elevations lower than 15$^\circ$.  \hl{The actual meteor arrived from a point in the sky well-inside the predicted region}.  In the V-C case, \hl{the ``forbidden region'' is smaller and two well-recognized spots at $A\sim 150^\circ, h\sim 10-40^\circ$ and $A\sim 40^\circ, h\sim 50-60^\circ$  concentrate the expected incoming directions}. The actual meteoroid came close to the first spot.

\begin{figure*}  
  \centering
  \vspace{0.2cm}
   \includegraphics[scale=0.35]{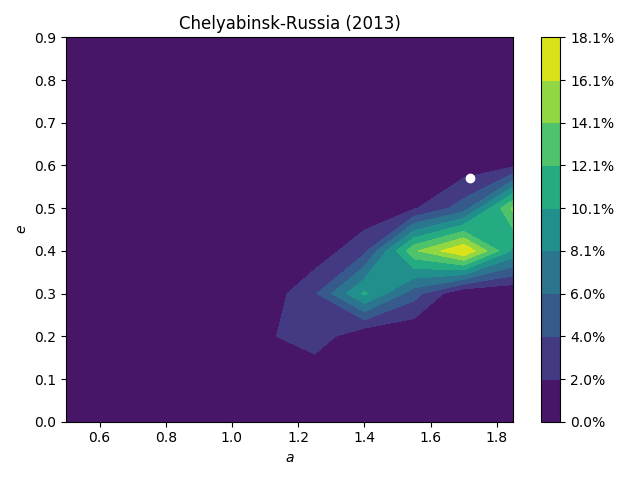}\hspace{0.2em}%
   \includegraphics[scale=0.35]{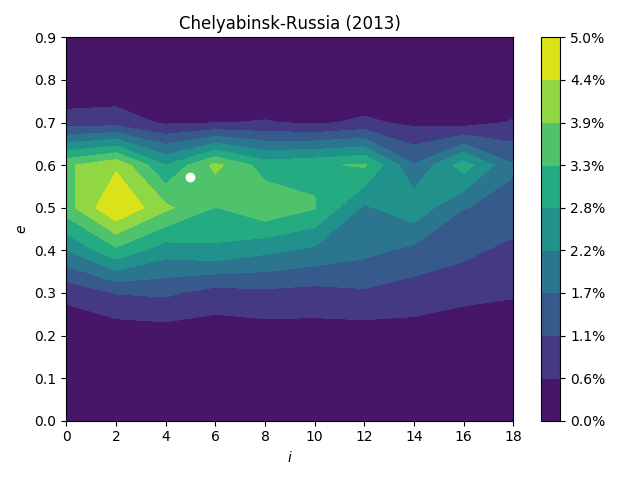}\hspace{0.2em}%
   \includegraphics[scale=0.35]{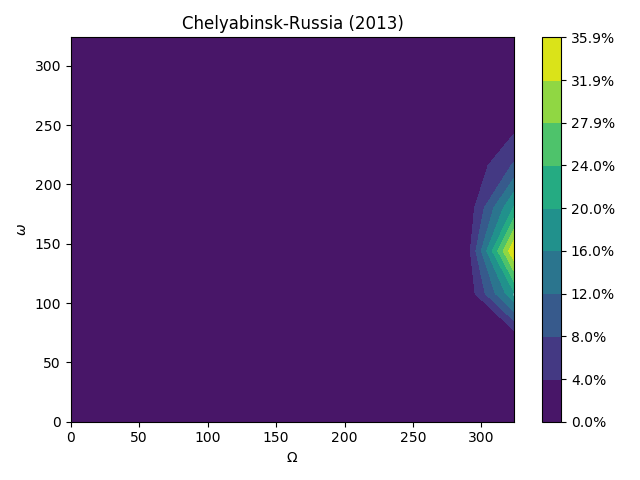}
   \\
   \includegraphics[scale=0.35]{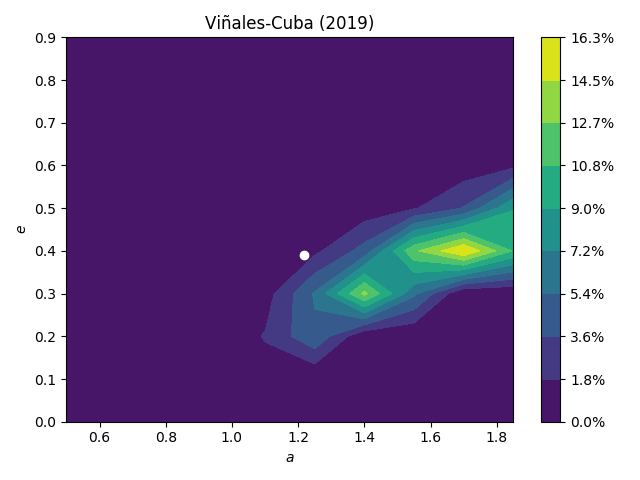} \hspace{0.2em}%
   \includegraphics[scale=0.35]{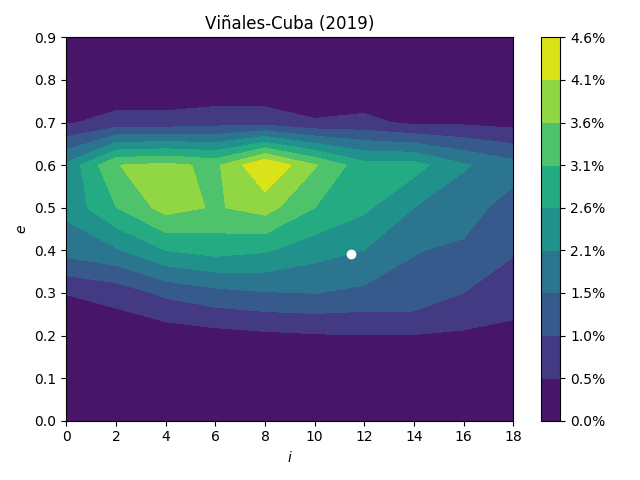} \hspace{0.2em}%
   \includegraphics[scale=0.35]{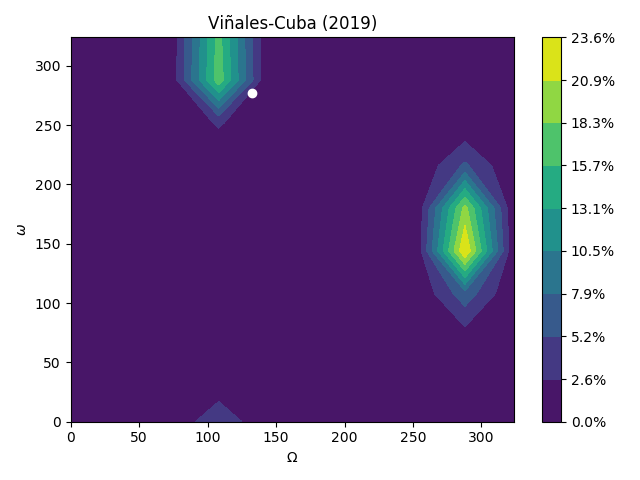}
\caption{Colourmaps show the regions in the configuration ($a$, $e$, $i$, $\omega$, $\Omega$) space more prone to impact the Earth at the time and place of Chelyabinsk (uppermost panels) and Cuban meteors (lowermost panels). }
\label{fig:orbital}
\end{figure*}

The \hl{heliocentric elements of the meteoroids} also show interesting coincidences with those predicted by GRT.  In Figure \ref{fig:orbital} we show bivariate marginal probability distribution for pairs of \hl{orbital} elements. For \hl{the C-R event}, GRT favor a low inclination meteoroid orbit with a moderate $e\sim0.4-0.6$ eccentricity \hl{which are close to the actual impactor orbit}. For the V-C event, GRT predicted \hl{a similar eccentricity but a slightly larger orbital inclination}. The actual body had $i\approx \iapprox^\circ, e\approx \eapprox$. The longitude of the ascending node $\Omega$ is easy to predict.  It will mostly depend on the ecliptic longitude of the Earth at the time of impact. However, the argument of the perihelion $\omega$, namely the orientation of the axes of the ellipse in the orbital plane, is not trivial to predict.  Still, GRT \hl{constrains} reasonably well the value of $\omega$ for both events.

\vspace{-0.4cm}

\section{Discussion}
\label{sec:discussion}

We started this letter by formulating a \hl{bold} question: {\it \mytitle} After the results presented in this work, \hl{which certainly does not constitute a statistically significant prove}, we propose \hl{also} a bold answer: {\it yes, we can predict them}. But this answer should not surprise us. As GRT intrinsically assumes, the NEOs population acts like a gravitational ``source of light'' that instead of photons sends particles towards the Earth.  Predicting the incoming impact direction is a simple consequence of knowing the properties of such body source.

\hl{The quality of GRT predictions depend on meteoroid population used for calculation ray probabilities. In this paper and for the sake of simplicity, we assumed that all potential impacting bodies have an orbital distribution similar to that of the {\it complete} NEO population ($\sim$20000 objects with absolute magnitud $H<20$ and diameters $D>500$ m). However, smaller and fainter objects, which are affected for instance by non-gravitational forces, may have different distributions (see eg. \citealt{Granvik2017}) or, as it is the case of centimetre-sized particles, they may have multiple and diverse sources (major meteor showers, sporadic meteor torus).  Other assumptions on the distribution of small and invisible impactors may be used to improve GRT following the recent results by \citet{Bouquet2014,Christou2014,Dmitriev2015,Granvik2016,Granvik2017}.}



\hl{Why did we focus on metre-sized meteoroids?. Three main reasons.  First, we can assume in a first approximation (as we have done in ZS2018) that they have a similar origin (and probably a similar distribution) as the population of larger NEOs. Even if this is not the case, performing simulations to study their true distribution is rather feasible.  Secondly, most of them are missing and still ``invisible'' to our surveys. And last but not least, they pose the worst risk to our civilization. Their impact rate is relatively high and the potential damaging effects, though small as compared with hundred-metre-sized objects, may have non-negligible economical impacts.} 

Is GRT perfect and their predictions entirely reliable? Probably not in its present form. As the differences between the predicted \hl{and the observed} impact conditions seems to reveal, the method can be still improved, \hl{not only with a better knowledge of the meteoroid orbital distribution but from a theoretical and numerical point of view}. However, it is hard for a single group of authors to achieve it.


Being able to constraint impact conditions will not solve all the risks imposed by small undetectable objects.  Still, it can help us to be prepared for future events, especially over populated areas.  With this work we also want to call the attention of governmental and no governmental institutions about the interest that theoretical research on impact risk assessment may have, and the potential that novel methods like GRT have at finding answer to what were considered unsolvable questions. 

\hl{Our results here relied in our own estimations of the V-C impact conditions. Although improved  conditions could be obtained combining all available information (satellite data, infrasound recordings, more and better footage, etc.)}, we are confident that the conclusions of this letter will not be substantially modified.

 


\vspace{-0.4cm}
\section*{Acknowledgements}

Most of the computations that made possible this work were performed with NASA NAIF SPICE Software (\citealt{Acton1996} and Jon D. Giorgini), {\tt Python 3.6} and their related tools and libraries, {\tt iPython} \citep{Perez2007}, {\tt Matplotlib} \citep{Hunter2007}, {\tt scipy} and {\tt numpy} \citep{Van2011}. \hl{We appreciate the insightful and constructive observations made by Maria Gritsevich during the review process.  We also thank her kind disposition.} We thank all the people in Cuba who shared their videos and footage on the internet, and ultimately allowed us to reconstruct the Cuba meteor impact conditions.  We are especially grateful to Rachel Cook who share its incredible video of the meteor from the Havana Harbor, which was instrumental for the successful reconstruction of the trajectory.  Karls Pe\~na and Rafael Colon of the Dominican Astronomical Society Astrodom (the Dominican Republic) helped us with the non-trivial task of finding information in Cuba about the meteor and the observation places. Thanks to them for this effort. 
\vspace{-0.3cm}
\bibliographystyle{mnras}
\bibliography{references.bib}

\clearpage
\appendix
\label{sec:supplementary}




\section{Supplementary material}

\noindent
{\Large \bf Impact conditions of the \hl{\Vinales}-Cuba meteor}

\subsection{The \hl{\Vinales}-Cuba meteor}

The \Vinales-Cuba meteor happened in February 1, 2019 around 18:17 UTC. It was witnessed by thousands in the island and several casual observers in south Florida (USA).  The meteor left a smoke trail and produced a sonic boom that recalled that of Chelyabinsk in 2013.  According to the NASA fireball database \footnote{\url{https://cneos.jpl.nasa.gov/fireballs/}} the meteor  released an estimated energy of $1.4$ kt of TNT. 

Here, we reconstruct the trajectory of the meteor in the atmosphere above Cuba and its orbit before the impact,  using for that purpose different footage originally discovered and obtained from {\tt YouTube}, {\tt Instagram} and {\tt Twitter}.  In particular we \hl{analysed} three videos taken at different {\it vantage points}, both in Cuba and USA.  Vantage points are separated by between 100 and 400 km ensuring a proper baseline for the reconstruction of the trajectory.

In Table \ref{tab:vantagepoints} we present the position of the vantage points and links to the corresponding footage we use in our reconstruction. 

\begin{table*}
\centering
\begin{tabular}{llllll}
\hline\hline
Location & long. (deg) & lat. (deg) & alt. (m) & Public & Alternative \\\hline
Havana Harbor, Cuba & -82.34433 & 23.13777 & 0 & \url{http://bit.ly/2GwIQqB} & \url{http://bit.ly/2UJ18c2} \\
Gull Wing Beach Resrot, FL USA & -81.90360 & 26.41887 & 0 & \url{http://bit.ly/2tgUU7q} & \url{http://bit.ly/2TGGp8A} \\
Alameda Pinar del Rio, Cuba & -83.69211 & 22.41455 & 0 & - & \url{http://bit.ly/2thGkMZ} \\\hline
\end{tabular}
\caption{Location of the vantages points where the footage used in this work were taken.  Public links are the original social network material.  Alternative are permanent link to the footage.}
\label{tab:vantagepoints}
\end{table*}

\subsubsection{Havana observations}
\label{sec:havana_observations}

Havana observations are based in a time-lapse, recorded on board of a cruiser in the Havana Harbor.  This video is the most complete and precise piece of footage publicly available about the event.  

The video was taken using a {\tt GoPro Hero 5} using the default time-lapse mode (0.5 seconds between frames). We achieved to recover 11 frames spanning a total of 5 seconds, where the meteor is clearly visible.  

\begin{figure}  
  \centering
  \includegraphics[width=0.50\textwidth]{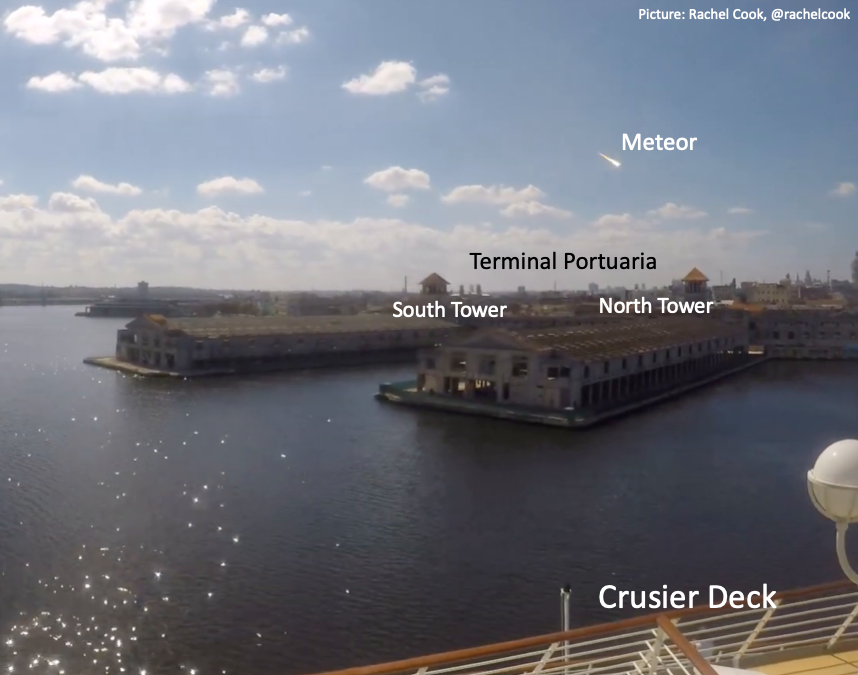}\\
  \includegraphics[width=0.50\textwidth]{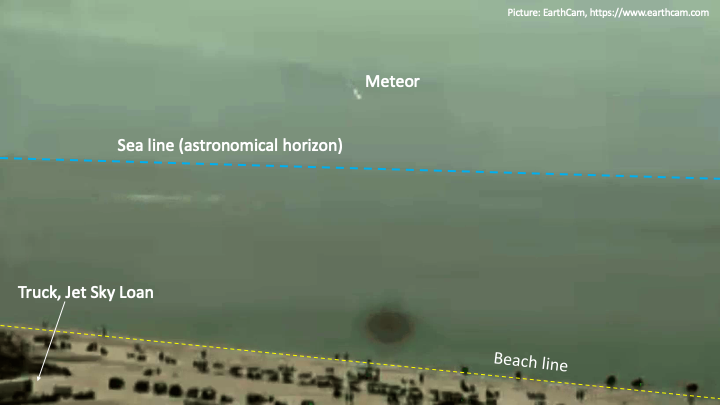}
\caption{Upper panel: a single frame of the video taken in the Havana Harbor showing the reference buildings used in this work to estimate azimuth and elevation of the meteor. Lowe panel: frame of a video taken in the Gull Wing Beach Resort (Floria, USA).\label{fig:havana_florida}}
\end{figure}

Azimuth and elevation of the meteor were estimated by first identifying several reference buildings on the horizon (see upper panel in Figure \ref{fig:havana_florida}).

\subsubsection{Florida observations}
\label{sec:florida_observations}

The second footage we use for our reconstruction is a video recorded by a camera of the {\tt Earth Cam} network\footnote{\url{https://www.earthcam.com/}} in the top of the {\it Gull Wing Beach Resort} at Ft. Myers Beach, Florida (USA).  A single frame of the video is shown in Figure \ref{fig:havana_florida}.

Although the quality of the video is poor (as compared with that of the Havana) and the weather near to the Horizons was partially cloudy, we get enough information from the video to estimate the (relative) azimuth and elevation of the meteor.

To obtain the angular scale of the picture, we use as a reference object a truck (or container) apparently used for a jet ski station.  The physical dimensions of the truck or container was obtained from a 3D model of the ``truck'' available in {\tt Google Earth}. Combining the estimated distance between the camera (which is located at the roof of the resort) and the truck, we calculate the field of view of each pixel in the image.  From that, we estimate the elevation of the meteor in each video frame.

Getting the azimuth of the meteor for this particular image was challenging. No fixed reference object could be identified along the beach.  Therefore, our azimuth measurements were only relative to an arbitrary direction on the horizon.  As we will see below, the precise azimuth of this reference direction was estimated a posteriori using the trajectory fitting procedure (Section \ref{sec:fit}).  

\subsubsection{Pinar del Rio observations}
\label{sec:pinar_observations}

The last footage we used for our reconstruction, was a street video showing the smoke trail left by the meteor. The video, recorded by a casual observer in the city of Pinar del Rio, 173 km to the west of Havana, was particularly well-suited for our purposes, in contrast with most videos recorded in the island, since it was plenty of reference objects able to provide us azimuths and the angular scale of the image. The video was apparently recorded a few minutes after the meteor. The site of the video was easily identified in \GoogleEarth. 

\begin{figure}  
\centering
\includegraphics[width=0.3\textwidth]{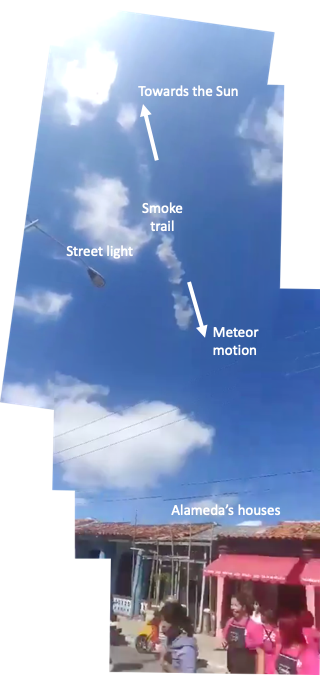}
\caption{A single frame of the video taken from Pinar del Rio (Cuba), showing the hallmarks used in this work to estimate azimuth and elevation of the meteor.}
\label{fig:pinar}
\end{figure}

We stitch together several of the frames in the video, showing the meteor trail and some reference objects, and the result is presented in Figure \ref{fig:pinar}.

In order to get the elevation and azimuth of several points of the smoke trail, we first need to obtain the angular scale of the picture.  For this purpose we use a lighting street pole located in the opposite side of the street.  According to a source in the island, the height of the lighting street poles in Cuba is Standard. We use this height and the distance of the observer to the pole to gauge the angular scale of the pictures.

The scarce information available in the pictures about the meteor, only allows us to estimate the position of just three points in the smoke trail:

\begin{itemize}
    \item the closest point to the top of a close lighting street pole (see Figure \ref{fig:pinar}). The azimuth of this point was assumed the same as that of the pole (measured using \GoogleEarth).
    \item The point in the smoke trail closer to the lamp in the street pole.  The elevation of this point was obtained after estimating the position of the zenith using the first measurement. A line drawn from the zenith to the lamp allows us to estimate the distance and hence the angle between the lamp and the smoke trail point. The azimuth of this line is the same as the azimuth of the vertical to the lamp as seen from the location of the observer.  The elevation of the lamp was estimated from the distance between the lamp and the observer. 
    \item The lowest point of the trail.  The azimuth of this point was estimated using as references the roof of houses below the trail. \GoogleEarth provided us the azimuth of those houses.  The elevation was estimated using the angular distance between the trail and the ground.
\end{itemize}  

The previous procedure has large uncertainties (several degrees in most of the cases) and it relies mostly on unknown properties (the height of the light pole, the azimuth of the houses, etc.)  Moreover, this procedure give us the observed coordinates of the smoke trail and not of the meteor itself.  Still, in the absence of other more reliable visual recording inside the island, the information provided by this images are priceless.

\bigskip

In Table \ref{tab:measurements} we summarize the observed coordinates (azimuth and elevation) of the meteor as estimated with the aforementioned procedures for the three vantage points.

\hl{In the case of Florida and Havana, we corrected the observed elevation by the effect of atmospheric refraction.  For this purpose, we initially use the formula by \citet{Bennett1982}:}

$$
h=h_a-\epsilon\cot\left(h_a+\frac{7.31}{h_a+4.44}\right)
$$

\hl{Where $h$ and $h_a$ are the true and the apparent elevations and $\epsilon=283 P/(273+T)$, with $P$ the atmospheric pressure in bars and $T$ is temperature in Celsius, is a factor taking into account the local atmospheric conditions.  This formula has been tested for navigation purposes and it is valid to all elevations.  Since the object was inside the atmosphere, we estimate that the atmospheric refraction from La Florida (400 km away from the impact site), was half of the total computed with the previous formula.  On the other hand, we estimated an atmospheric refraction of 1/3 of the total for the observations done from Cuba ($\sim$200 km away from the meteor trajectory).}

\begin{table}
\centering
\begin{tabular}{llll}
\hline\hline
Point & Az. (deg) & h (deg) & t (s)\\
\hline\multicolumn{4}{c}{Havana harbor}\\\hline
1 & 226.74 $\pm$ 0.9 & 18.6 $\pm$ 0.07 & 0.00 \\
2 & 228.22 $\pm$ 0.9 & 17.7 $\pm$ 0.07 & 0.50 \\
3 & 229.77 $\pm$ 0.9 & 17.0 $\pm$ 0.07 & 1.00 \\
4 & 231.51 $\pm$ 0.9 & 16.0 $\pm$ 0.07 & 1.50 \\
5 & 233.13 $\pm$ 0.9 & 15.0 $\pm$ 0.07 & 2.00 \\
6 & 235.00 $\pm$ 0.9 & 14.2 $\pm$ 0.07 & 2.50 \\
7 & 236.87 $\pm$ 0.9 & 13.0 $\pm$ 0.07 & 3.00 \\
8 & 238.82 $\pm$ 0.9 & 11.8 $\pm$ 0.5 & 3.50 \\
9 & 240.89 $\pm$ 0.9 & 10.6 $\pm$ 0.5 & 4.00 \\
10 & 243.00 $\pm$ 0.9 & 9.2 $\pm$ 0.5 & 4.50 \\
11 & 244.96 $\pm$ 0.9 & 7.9 $\pm$ 0.5 & 5.00 \\
\hline\multicolumn{4}{c}{GullWing Beach Resort (Florida)}\\\hline
1 & $A_{\rm ref}$+10.1 $\pm$ 0.1 & 2.4 $\pm$ 0.3 & 0.44 \\
2 & $A_{\rm ref}$+10.2 $\pm$ 0.1 & 2.2 $\pm$ 0.3 & 0.50 \\
3 & $A_{\rm ref}$+10.5 $\pm$ 0.1 & 1.8 $\pm$ 0.3 & 0.81 \\
4 & $A_{\rm ref}$+10.8 $\pm$ 0.1 & 1.5 $\pm$ 0.3 & 0.94 \\
5 & $A_{\rm ref}$+11.2 $\pm$ 0.1 & 0.9 $\pm$ 0.3 & 1.94 \\
6 & $A_{\rm ref}$+11.5 $\pm$ 0.1 & 0.6 $\pm$ 0.3 & 2.31 \\
\hline\multicolumn{4}{c}{Pinar del Rio (Cuba)}\\\hline
1 & 287.61 $\pm$ 0.1 & 71.4 $\pm$ 5 & - \\
2 & 305.01 $\pm$ 5 & 62.8 $\pm$ 5 & - \\
3 & 318.75 $\pm$ 5 & 57.1 $\pm$ 5 & - \\
\hline\hline
\multicolumn{4}{l}{\footnotesize*$A_{\rm ref}$ is obtained in the fitting procedure.}\\
\end{tabular}
\caption{Observed trajectory of the meteor in the sky as estimated in this work for three different vantage points.\label{tab:measurements}}
\end{table}

\subsection{Trajectory fitting}
\label{sec:fit}

One of the main limitations of working with public or amateur footage is the lack of a proper synchronization among videos. In only two of them (the Havana and Florida videos) we have information about the time of each frame relative to the beginning of the video (last column in Table \ref{tab:measurements}).  No absolute timing was available.  

For solving this inconvenience we developed in Zuluaga et al. (2013) a numerical procedure intended to fit a meteor trajectory, starting only with the measurements of elevations and azimuths. We call it the ``Altazimuth-footprint method''. 

In contrast with the original version of the method, the improved version we use here, involved both, errors in elevation and azimuth to compute the altazimuth-footprint statistics:

\beq
\label{eq:altaz}
\mathcal{A}=\sqrt{
\sum_{j=1}^{n_v}\sum_{i=1}^{n_j} 
\left[
\frac{(A_{ji}-A_{tji})^2}{\Delta A_{ji}^2}+
\frac{(h_{ji}-h_{tji})^2}{\Delta h_{ji}^2}
\right]
}
\eeq

Here, $n_v$ is the number of vantage points, $n_j$ is the number of observations in the $j$th vantage point, $(A_{ji},h_{ji})$ are the horizontal coordinates of the $i$th point in the $j$th vantage point, and $(A_{tji},h_{tji})$ are the azimuth and elevation of the closest point in the theoretical trajectory respectively.

The theoretical trajectory depends on four quantities: the projected impact site location (lon$_{\rm imp}$,lat$_{\rm imp}$) (alt$_{\rm imp}$=0), the radiant azimuth $A_{\rm rad}$ and the radiant elevation $h_{\rm rad}$ as seen from the projected impact site. 

In order to find the best-fit value of those parameters we minimize the statistics ${\cal A}$. Due to the particularities of the Florida footage, we add a fifth parameter to the minimization procedure, namely the reference Azimuth $A_{\rm ref}$ at the Gull Wing Beach vantage point (see Table \ref{tab:measurements}).

The result of the fitting procedure is presented in Table \ref{tab:fit}. Using the best-fit trajectory, we also compute the heliocentric orbit of the meteoroid 1 year before the impact. The resulting classical elements are also shown in this Table.

\begin{table}
\centering
\begin{tabular}{ccc}
\hline\hline
Parameter & Best fit value\\
\hline
\multicolumn{2}{c}{Atmospheric trajectory}\\
\hline
Lon. impact (deg) & \hl{-83.8037 $\pm$ 0.002} \\
Lat. impact (deg) & \hl{+22.8820 $\pm$ 0.001} \\
$A_{\rm rad}$ (deg) & \hl{178.9 $\pm$ 0.2} \\
$h_{\rm rad}$ (deg) & \hl{31.8 $\pm$ 0.1} \\
$A_{\rm ref}$ (deg) & \hl{193.1} \\
$\langle v_{\rm imp}\rangle$ (km/s) & $16.9^{+0.15}_{-0.14}$ \\
\hline
\multicolumn{2}{c}{Orbit}\\
\hline
$a$ (AU) & ${1.217}^{+0.003}_{-0.004}$\\
$q$ (AU) & ${0.740}^{+0.004}_{-0.005}$\\
$e$ & ${0.391}^{+0.005}_{-0.007}$\\
$i$ (deg) & ${11.471}^{+0.030}_{-0.038}$\\
$\Omega$ (deg) & ${132.281}^{+0.002}_{-0.002}$\\
$\omega$ (deg) & ${276.975}^{+0.434}_{-0.507}$\\
$P$ (yr) & ${1.342}^{+0.006}_{-0.007}$\\
$T_p$ & ${2.811}^{+0.005}_{-0.005}$\\

\hline
\end{tabular}
\caption{Best-fit values of the atmospheric trajectory and the asymptotic orbit of the meteoroid for the Cuba impact as estimated in this work.}
\label{tab:fit}
\end{table}

\begin{figure*}  
  \centering
  \includegraphics[width=0.50\textwidth]{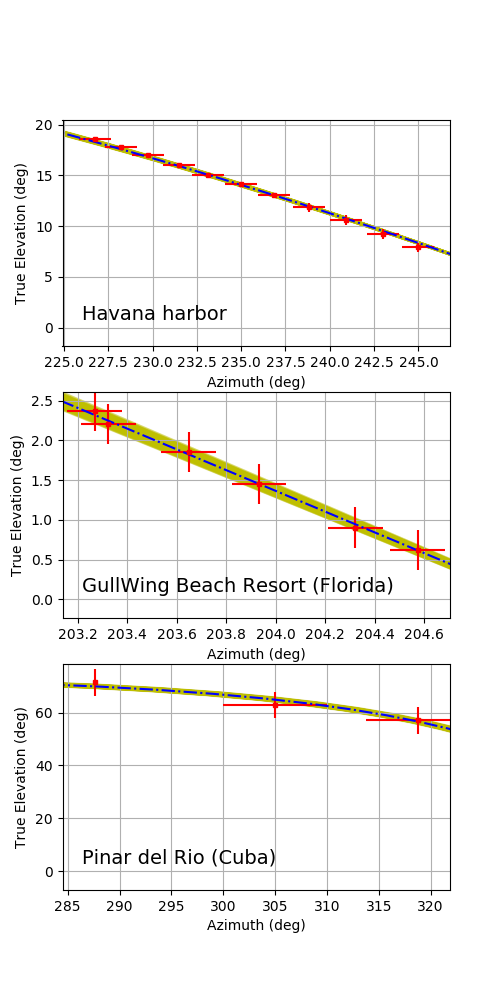}
\caption{Expected azimuth and elevation of the theoretical trajectory (dashed-dotted line) as observed from the vantage points, and the actual measurements estimated in this work (crosses).  In the three cases the horizontal and vertical ranges have been adjusted to the ranges of the available data.  The shaded region correspond to the theoretical trajectories calculated for the range of impact conditions statistically consistent with the data at 95\% C.L.\label{fig:fit}}
\end{figure*}

A plot with a comparison between the observed meteor trajectory and the best-fit Altazimuthal footprint, is presented in Figure \ref{fig:fit}.

We found that in order to explain our three visual observations, the object should come almost exactly from the south, \hl{travelling} in an inclined \hl{$h_{\rm rad}=\hrad^\circ$} trajectory towards the north coast of Cuba.  According to the observations in the Havana Harbor the contact point of the object with the atmosphere happened at \hl{\maxheight} km above the ocean to the south of San Felipe Cays.  The brightest point in the video corresponds to a point in the trajectory \hl{\maxbright} km above the surface.  

On the other hand, according to our estimations, the smoke trail seen in the Pinar del Rio video, corresponds to a short segment of the trajectory between \hl{\pinarhmax} km and \hl{\pinarhmin} km.

Using our fitted trajectory and the very precise footage taken at the Havana harbor, we can estimate the distance \hl{travelled} by the meteoroid in the atmosphere as a function of time.  A linear regression of the resulting distances, provide us the average speed of the meteoroid during the airburst.

\begin{figure*}  
  \centering
  \includegraphics[width=0.45\textwidth]{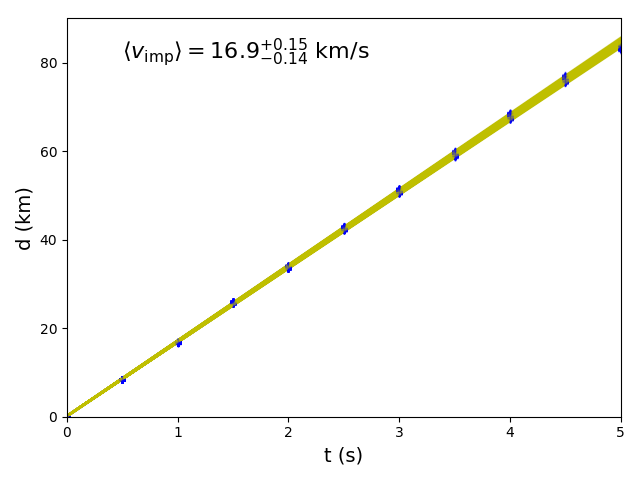}
\label{fig:speed}
\caption{Distance travelled by the meteor using trajectories consistent with the available data (shadow region), and the Havana Harbor observations (black dots).  The average impact speed $\langle\vimp\rangle$ and its error, are the slope of the lines in the shaded region.}
\end{figure*}


\subsection{Discussion}

The estimated impact conditions, presented here, depends on several assumptions and on the validity of the geometrical procedures used to measure azimuths and elevations over the pictures.  


The most uncertain measurements are those taken at Pinar del Rio.  Since the footage there does not include the visible meteor but the smoke trail, it could happen that the cloud be displaced from the actual meteor trajectory by several degrees. In particular the meteor could happen to the west of the cloud observed position.  In this case the trajectory will be farther away from the Havana Harbor and hence the estimation of the distance \hl{travelled} by the meteoroid, and hence the impact speed, will also be affected.  

We have played around with the observed coordinates of the Pinar del Rio observations, modifying elevations and azimuths to simulate the effect of the wind.  The errors reported in Table \ref{tab:fit} reflect those uncertainties.

In order to estimate the errors in the orbital elements, we create 500 test particles having random impact conditions within the intervals defined by the errors in the atmospheric trajectory section of Table \ref{tab:fit}.  The errors reported in the orbital elements are the result of this simulations.

\subsection{Reproducibility}

In order to make all our results reproducible, we provide all the input data, simulation results, spreadsheets and crude videos and images, in a public {\tt GitHub} repository: \url{http://github.com/seap-udea/MeteorTrajectories.git}.  The GRT method was implemented in a {\tt C$++$/python} package  also publicly available in the {\tt GitHub} repository \url{http://github.com/seap-udea/GravRay.git} (branch {\tt MoonImpact}).

Any suggestion, observations or corrections will be greatly appreciated.

\bsp
\label{lastpage}
\end{document}